%% file: bare_conf.tex
\documentclass[conference]{IEEEtran}
\ifCLASSINFOpdf
\else
\fi
\usepackage{cite}
\usepackage{bm}
\usepackage{float}
\usepackage{amsmath,amssymb,amsfonts}
\usepackage{mathtools, nccmath}
\usepackage{graphicx}
\usepackage{textcomp}
\usepackage{xcolor}
\usepackage[font=small]{caption}
\usepackage{subcaption}
\usepackage{nicematrix}
\usepackage[shortlabels,inline]{enumitem}
\usepackage{comment}
\usepackage{cuted}
\usepackage[para,flushleft]{threeparttable}
\usepackage{adjustbox}
\usepackage{booktabs}
\usepackage{color, colortbl}
\usepackage{soul}
\usepackage{multirow}
\usepackage{tikz}
\usepackage{pgfplots}
\usepackage{algorithm}
\usepackage{siunitx}
\usepackage[noend]{algpseudocode}%
\usepackage[hidelinks]{hyperref}
\usepackage{arydshln}
\pgfplotsset{width=10cm,compat=1.9}
\definecolor{Gray}{gray}{0.9}
\usepackage{svg}

\newcommand{\ceil}[1]{\lceil #1 \rceil}

\newcommand{\rt}[1]{\hat{#1}}

\makeatletter
\newcommand{\linebreakand}{%
  \end{@IEEEauthorhalign}
  \hfill\mbox{}\par
  \mbox{}\hfill\begin{@IEEEauthorhalign}
}
\makeatother

\begin{document}

\setlength{\abovedisplayskip}{3pt}
\setlength{\belowdisplayskip}{3pt}

\title{HARFLOW3D: A Latency-Oriented 3D-CNN Accelerator Toolflow for HAR on FPGA Devices}

\author{
    \IEEEauthorblockN{Petros Toupas\IEEEauthorrefmark{1}$^{1,2}$}
    \and
    \IEEEauthorblockN{Alexander Montgomerie-Corcoran\IEEEauthorrefmark{1}$^{1}$}
    \and
    \IEEEauthorblockN{Christos-Savvas Bouganis$^1$}
    \and
    \IEEEauthorblockN{Dimitrios Tzovaras$^2$}
    \linebreakand
    \IEEEauthorblockA{
    $^1$
    Dpt. of Electrical and Electronic Engineering\\
    Imperial College London\\
    Email: \{p.toupas21, \\alexander.montgomerie-corcoran15,\\ christos-savvas.bouganis\}@imperial.ac.uk}
    \and
    \IEEEauthorblockA{
    $^2$
    Information Technologies Institute\\
    Centre of Research and Technology Hellas\\
    Email: \{ptoupas,dimitrios.tzovaras\}@iti.gr}
}

\maketitle

\begingroup\renewcommand\thefootnote{\IEEEauthorrefmark{1}}
\footnotetext{\textit{Equal contribution}}
\endgroup

\begin{abstract}
For Human Action Recognition tasks (HAR), 3D Convolutional Neural Networks have proven to be highly effective, achieving state-of-the-art results. This study introduces a novel streaming architecture-based toolflow for mapping such models onto FPGAs considering the model's inherent characteristics and the features of the targeted FPGA device. The HARFLOW3D toolflow takes as input a 3D CNN in ONNX format and a description of the FPGA characteristics, generating a design that minimises the latency of the computation. The toolflow is comprised of a number of parts, including 
\begin{enumerate*}[label=(\roman*)]
    \item a 3D CNN parser,
    \item a performance and resource model,
    \item a scheduling algorithm for executing 3D models on the generated hardware,
    \item a resource-aware optimisation engine tailored for 3D models,
    \item an automated mapping to synthesizable code for FPGAs
\end{enumerate*}.
The ability of the toolflow to support a broad range of models and devices is shown through a number of experiments on various 3D CNN and FPGA system pairs.
Furthermore, the toolflow has produced high-performing results for 3D CNN models that have not been mapped to FPGAs before, demonstrating the potential of FPGA-based systems in this space.
Overall, HARFLOW3D has demonstrated its ability to deliver competitive latency compared to a range of state-of-the-art hand-tuned approaches, being able to achieve up to 5$\times$ better performance compared to some of the existing works.
The tool is available at \underline{\url{https://github.com/ICIdsl/harflow3d}.}
\vspace{0.2cm}
\end{abstract}


\IEEEpeerreviewmaketitle

\input{Introduction/introduction}

\input{Background/background}

\input{Architecture/architecture.tex}

\input{Modelling/modelling}

\input{DesignSpaceExploration/dse}

\input{Validation/validation}

\input{Evaluation/evaluation}
\input{Conclusion/conclusion}

\section*{Acknowledgment}
For the purpose of open access, the authors have applied a Creative Commons Attribution (CC BY) license to any Accepted Manuscript version arising.

\bibliographystyle{IEEEtran}
\bibliography{references}

\end{document}

%% file: Introduction/introduction.tex
\section{Introduction} \label{introduction}

The growing focus on video-related applications such as video surveillance, autonomous driving, and patient monitoring has necessitated the development of algorithms that integrate and take into account the temporal domain.
3D CNNs, which are often employed to deal with video and volumetric data, augment their learning capacity by extracting input features related to this additional dimension.
Due to the temporal dimension, 3D CNNs often have larger computational and memory requirements compared to 2D CNNs.
Particularly, 3D CNNs have exhibited high performance in the task of HAR, enabling the interpretation of human motion across video frames and the detection of various activities without the need for specialised time domain approaches (e.g., LSTMs).
Whilst vision transformers have recently attained state-of-the-art accuracy, their operation requires orders of magnitude more GFLOPs comparatively.

Devices such as GPUs, FPGAs, and ASICs have been utilised to address the high processing requirements of 3D CNNs and deliver high-performance systems.
FPGAs are particularly attractive as an acceleration platform since they are more flexible than ASICs and more energy efficient than GPUs.
The rapid growth and increasing complexity of 3D CNN model designs necessitate high-quality hardware designs that support short design cycles for new 3D CNN model specifications.
The goal of this work is to provide an automated way for deploying 3D CNN models onto FPGA systems, with an emphasis on minimising the execution latency. The diversity of the supported models and devices makes it suitable for a number of applications and enables users to select the most appropriate 3D CNN model and device according to their unique demands and budget.

\begin{figure}
    \centering
    \includegraphics[width=\columnwidth, keepaspectratio]{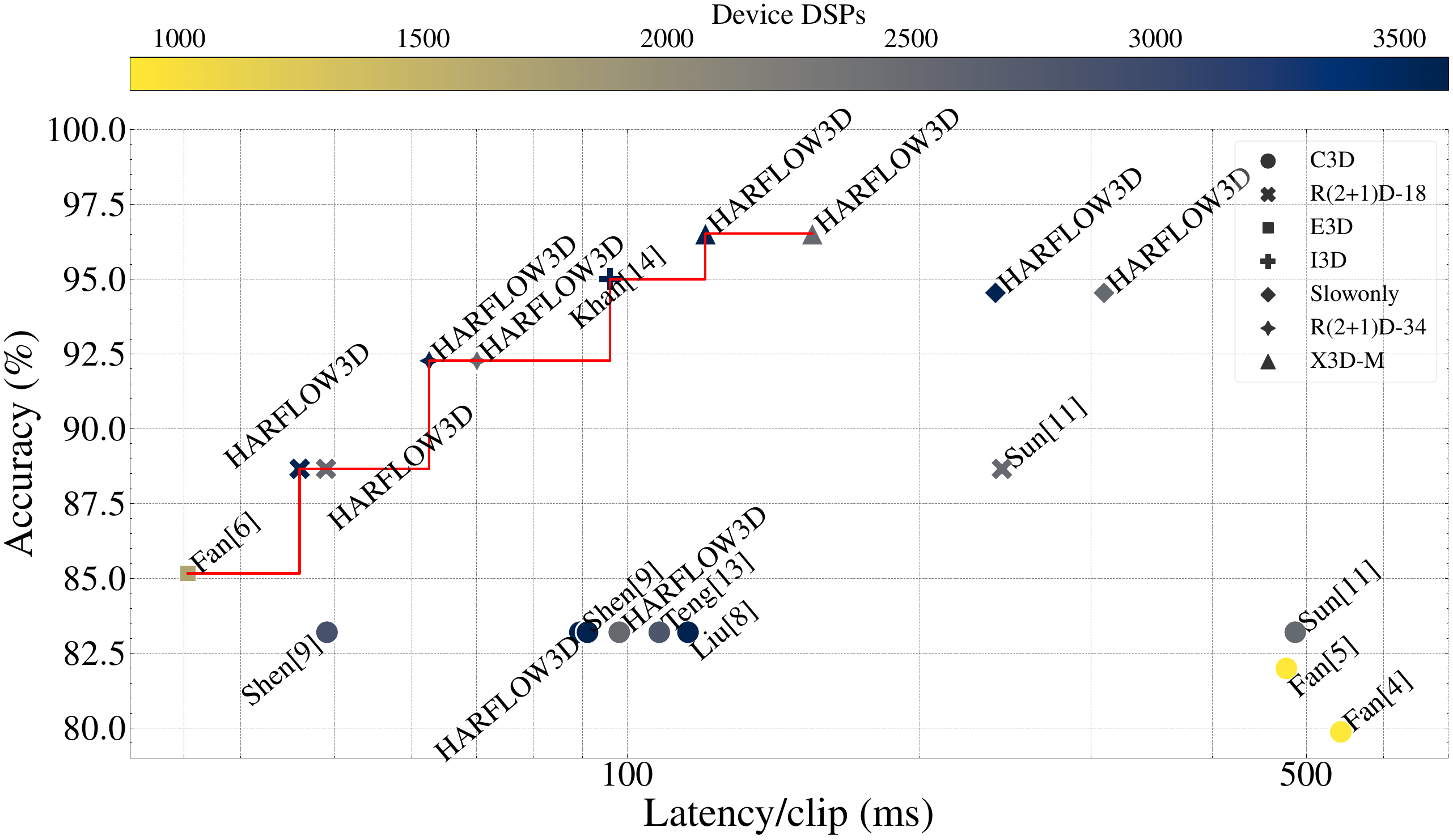}
    \caption{Pareto front on 3D CNNs: Latency over Accuracy. Designs produced by the proposed HARFLOW3D toolflow dominate the pareto front.}
    \label{fig:perf_accuracy_pareto}
\end{figure}

3D CNNs have been studied and developed for quite some time, and the topic of HAR is gaining attention year by year. A few studies have focused on mapping 3D CNNs to FPGAs, but the vast majority of them propose hand-tuned hardware architectures for specific 3D CNN models. While FPGA toolflows for 2D CNNs are well-researched \cite{Chen2020ANetworks, Venieris2018a}, there is an absence of 3D CNN FPGA toolflows. The distinct characteristics of 3D CNNs, such as their large workloads and significantly increased memory and resource requirements, require such toolflows.

Figure \ref{fig:perf_accuracy_pareto} displays the pareto-front both for prior works as well as the proposed toolflow, showing their achieved accuracy and latency.
HARFLOW3D designs account for most of the points on the pareto-front, demonstrating the toolflow's ability to generate pareto-optimal designs for a variety of 3D CNN models on HAR.
The pareto-optimal relationship between accuracy and latency is an extremely desirable feature for a toolflow, as it allows a designer to trade-off their model's accuracy for greater performance in a fine-grain manner.

The key contributions of this paper are the following:
\begin{itemize}
    \item Introduction of HARFLOW3D, the first 3D CNN to FPGA toolflow, which supports a variety of models and devices, achieving competitive results compared to prior hand-tuned works.
    \item An optimisation strategy accompanied by a set of transformations, providing tailored designs based on the characteristics of each 3D CNN model layer.
    \item A set of highly-tuned parameterized building blocks that support runtime parameterization, allowing for significantly lower latency over the non-parameterized equivalent.
    \item A rich evaluation with experimental results across multiple devices and multiple models, including state-of-the-art 3D CNN HAR models that have not been addressed before, setting the landscape for FPGA-based HAR model computation.
\end{itemize}

%% file: Background/background.tex
\section{Related Work} \label{background}
Whilst 3D CNNs have been around for a while, there have been few studies that focus on accelerating these networks on FPGAs. The majority of these studies have focused on older 3D CNNs such as the C3D\cite{Ji2013} model, whose accuracy falls short of state-of-the-art models. Fan et al. \cite{Fan2017, Fan2018, Fan2019} have released a series of publications about accelerating 3D CNNs for human action recognition using FPGAs. In their first work\cite{Fan2017}, they introduced the F-C3D hardware architecture for accelerating the C3D model, which is capable of supporting multiple 3D convolutional layers and includes solutions for resolving some of the challenges posed by 3D CNNs, such as higher processing and memory demands. Additionally, they demonstrated the portability of their design to other FPGA devices. In a following publication\cite{Fan2018}, they presented an analytical model and a tool for optimising the hardware architecture based on device specifications, accuracy requirements, and the usage of block floating point arithmetic precision to minimise accuracy loss. Their evaluation of the design was likewise performed on the C3D model. In their most recent publication\cite{Fan2019}, they proposed E3DNet, an effective 3D CNN based on the proposed 3D-1 bottleneck building block. The F-E3D hardware implementation of E3DNet achieves a real-time execution time of 35.3 milliseconds per frame and scores 85.1\% accuracy on the UCF101\cite{Soomro2012UCF101:Wild} benchmark.

Based on the similarities between the 2D and 3D convolution computing patterns, Liu et al.\cite{Liu2019AFpgas} presented a hardware design to accelerate both 2D and 3D CNNs. They sought to turn CNN convolutions into matrix multiplication operations and concentrated on minimising memory usage to overcome the challenges associated with replicating feature maps. Applying an analytical model, they configured the accelerators for maximum resource utilisation and evaluated their design using the C3D model. Shen et al.\cite{Shen2018} developed a template-based architecture based on the Winograd transform~\cite{Winograd1980ArithmeticComputations} that is capable of handling both 2D and 3D CNNs. In addition, they developed an analytical method for quickly exploring the design space for mapping 2D and 3D CNNs onto FPGA accelerators and validated their design with the C3D model. Sun et al.\cite{Sun20203DPruning} applied weight pruning to the C3D and R(2+1)\cite{Tran2018} 3D CNN architectures using a blockwise approach. Their hardware architecture, which is based on the Alternating Direction Method of Multipliers (ADMM), enables the acceleration of 3D CNNs with minimal accuracy loss as compared to their unpruned counterparts.

Teng et al.\cite{Teng2020ExplorationAccelerator} presented a design space exploration strategy for optimising memory access in 3D CNN models accelerated on FPGAs. The authors proposed a non-overlapping data tiling method for off-chip memory access and explored on-chip data reuse using different loop ordering strategies. They further proposed a hardware architecture that can support these strategies. Their experiments showed that the proposed approach on the C3D model achieved state-of-the-art performance compared to prior FPGA implementations at that time. Khan et al.\cite{Khan2023TowardsNetworks} investigated various 3D CNN design parameters for resource-limited platforms, focusing on the I3D model, a 70-layer deep network for video action recognition. They adjusted the feature-map word lengths and weights in a pre-trained model, which reduced its complexity without affecting its accuracy. They proposed a data tiling technique that utilises all four dimensions of video data and improves memory bandwidth while reducing DRAM accesses. Based on these optimisations, the proposed FPGA accelerator achieves 684 GOPs/s for 32-bit floating point and 1.29 TOPs/s for 8-bit integer implementations with a 2\% accuracy drop.

The majority of research has been largely focused on the C3D~\cite{Ji2013} model for HAR, which was introduced in 2013. The model's architecture is rather simple, consisting of only 8 convolutional layers, while it performs poorly in terms of accuracy when compared to recent SoA models in HAR (85.2\% in UCF101 vs. the current SoA's 98.6\%). In terms of design complexity, an analogy can be made to AlexNet \cite{KrizhevskyImageNetNetworks}, but in three-dimensional space. Since the aforementioned approaches are mostly focused on the design of the specific C3D model, it is not clear how they can be extended, evaluated, or applied to the more complicated networks of modern state-of-the-art HAR models. The proposed toolflow addresses this by supporting a broad variety of recent 3D CNN designs, such as X3D~\cite{Feichtenhofer2020} and Slowonly~\cite{Feichtenhofer2019} among others, which include more complex ResNet3D-wise architectures as well as older models like C3D for direct comparison with prior studies.

%% file: Architecture/architecture.tex
\section{Proposed Architecture} \label{sec:architecture}

This section outlines the basics of the hardware-level architecture of the toolflow's generated designs. 
The suggested toolflow adheres to the same streaming architecture principles as fpgaConvNet\cite{Venieris2019}, which is based on the Synchronous Data-Flow (SDF) computation model. 

\subsection{Neural Network Model Parser}

To facilitate the process of incorporating various neural network (NN) models into the toolflow, a dedicated NN model parser has been developed. This parser is designed to read and map NN models in the ONNX format, a standardised format supported by many deep learning frameworks such as PyTorch and TensorFlow, into the format required by the toolflow. The 3D CNN model, can be described as a Directed Acyclic Graph (DAG), which is denoted as $M = \{l_1, ..., l_L\}$, where $l_i$ is the $i^{th}$ layer within the set of model layers $L$ and is expressed as an execution node of $M$.
This is then translated into a Synchronous Data-Flow Graph (SDFG) which is also directed and acyclic, denoted as $G$ with $N$ computation (or hardware) nodes, where $G = \{n_1, ..., n_N\}$.
The core concept of synchronous dataflow modelling is that each node fires whenever data is available at its inputs, resulting in a paradigm of data-driven execution. 
This format is compatible with the rest of the toolflow's tools, such as the latency optimiser and the toolflow's resource and performance models.

\subsection{Building Blocks Description} \label{architecture:bblocks}

Each layer of a NN model accepts input data to be processed and returns output data once its operation has been completed. These inputs and outputs are described as feature-maps in and out respectively. The maximum feature-map dimensions supported for a given hardware node $n$ are described as below.
\begin{align*}
    \bm{S_{n}^{in}}  &= \{ H_{n}^{in},  W_{n}^{in}, D_{n}^{in},  C_{n}^{in}  \} \\
    \bm{S_{n}^{out}} &= \{ H_{n}^{out}, W_{n}^{out}, D_{n}^{out}, C_{n}^{out} \}
\end{align*}
where $H_{n}^{in/out}$, $W_{n}^{in/out}$, $D_{n}^{in/out}$, and $C_{n}^{in/out}$ are the spatial dimensions (Height, Width), followed by the temporal dimension (Depth), followed by the number of Channels for the input and output respectively. The size of the feature-map in regards to the number of elements is referred to as $|\bm{S}|$. 
The $\bm{S}_n^{in}$, and $\bm{S}_n^{out}$ parameters exist for all of the layers as part of their parameter space definition. 
Alongside functional parameters, the hardware accepts different fixed-point precisions at compile-time.
The rest of the parameters of the layers, and their intermediate representation definitions are detailed in Table \ref{tab:building_blocks}, and a description is provided below.
The runtime parameters are differentiated from compile-time parameters using the $\wedge$ symbol above a parameter. Computation node parameters are subscripted with $n$, and execution node parameters with $l$.

\begin{table}
    \centering
    \caption{Compile time parameters for node $n$ in the hardware graph $G$, for each layer type.}
    \label{tab:building_blocks}
    \begin{tabular}{|c|l|}
    \hline
    \multicolumn{2}{|l|}{\textbf{Convolution}} \\ \hline
        $F_n$       & Number of filters (output channel dimension) \\ \hline
        $\bm{K}_n$  & 3D kernel size ($K_n^D$, $K_n^H$, $K_n^W$) \\ \hline
        $\bm{J}_n$  & 3D stride ($J_n^D$, $J_n^H$, $J_n^W$,) \\ \hline
        $\bm{P}_n$  & 3D Padding ($P_n^{D_s}$, $P_n^{D_e}$, $P_n^{H_s}$, $P_n^{H_e}$, $P_n^{W_s}$, $P_n^{W_e}$) \\ \hline
        $Gr_n$      & Grouping along the channel dimension \\ \hline
        $c_n^{in}$  & parallel streams in \\ \hline
        $c_n^{out}$ & parallel streams out \\ \hline
        $f_n$       & Vector dot product folding \\ \hline
        \multicolumn{2}{|l|}{\textbf{Fully Connected}} \\ \hline
        $F_n$       & Number of filters (output channel dimension) \\ \hline
        $c_n^{in}$  & Number of parallel streams in \\ \hline
        $c_n^{out}$ & Number of parallel streams out \\ \hline
        \multicolumn{2}{|l|}{\textbf{Pooling}} \\ \hline
        $T_n$       & Type of activation \\ \hline 
        $\bm{K}_n$  & 3D kernel size ($K_n^D$, $K_n^H$, $K_n^W$) \\ \hline
        $\bm{J}_n$  & 3D stride ($J_n^D$, $J_n^H$, $J_n^W$,) \\ \hline
        $\bm{P}_n$  & 3D Padding ($P_n^{D_s}$, $P_n^{D_e}$, $P_n^{H_s}$, $P_n^{H_e}$, $P_n^{W_s}$, $P_n^{W_e}$) \\ \hline
        $c_n$       & Number of parallel streams in \& out \\ \hline
        \multicolumn{2}{|l|}{\textbf{Activation}} \\ \hline
        $T_n$       & Type of activation \\ \hline
        $c_n$       & Number of parallel streams in \& out \\ \hline
        \multicolumn{2}{|l|}{\textbf{Global Average Pooling}} \\ \hline
        $c_n$       & Number of parallel streams in \& out \\ \hline
        \multicolumn{2}{|l|}{\textbf{Element-Wise}} \\ \hline
        $T_n$       & Type of element-wise operation \\ \hline
        $B_n$       & Mode of operation (default or broadcast) \\ \hline
        $c_n$       & Number of parallel streams in \& out \\ \hline
    \end{tabular}
\end{table}

\begin{itemize}
    \item \textbf{Convolution 3D} Due to the necessity to support a range of 3D CNN models, the toolflow's convolution building block is designed to accommodate the following types of convolution operation:
    \begin{enumerate*}[(a),leftmargin=+.4in]
        \item Full convolution $K^D \times K^H \times K^W$
        \item Spatial convolution $1 \times K^H \times K^W$
        \item Temporal convolution $K^D \times 1 \times 1$
        \item Depth-wise convolution
        \item Point-wise convolution
    \end{enumerate*}.
    The computation node is characterised as the following tuple:
    $$
        \Gamma = \{ \rt{\bm{S}}^{in}, \rt{\bm{S}}^{out}, \rt{\bm{K}}, \rt{\bm{J}}, 
        \rt{\bm{P}}, \rt{Gr}, \rt{c}^{in}, \rt{c}^{out}, \rt{f} \}
    $$

        
    \item \textbf{3D Pooling} For 3D pooling layers, the toolflow supports both maximum and average pooling which can be chosen at runtime by the enumerated parameter $T$. 
        The computation node is characterised as the following tuple:
        $$
            \Gamma = \{ \rt{\bm{S}}^{in}, \rt{\bm{S}}^{out}, \rt{\bm{K}}, \rt{\bm{J}}, \rt{\bm{P}}, \rt{T}, \rt{c} \}
        $$ 
    \item \textbf{Global Pooling, Activation \& Element-Wise 3D} 
        Although Global Pooling is a special case of the regular Pooling layer, the hardware for it is optimised for this case.
        The supported activation functions of the activation layer are the following: \begin{enumerate*}[(a),leftmargin=+.4in]
            \item ReLU activation,
            \item Sigmoid activation,
            \item Swish activation ($ y = x * sigmoid(x) $)
        \end{enumerate*}
        , where the type of activation is given by the parameter $T$.
        The runtime choice for broadcasting is defined as $B$, which is either true or false.
        The type of element-wise operation is given by the parameter $T$. 
        The computation node is characterised as the following tuple:
        $$
            \Gamma = \{ \rt{\bm{S}}^{in}, \rt{\bm{S}}^{out}, \rt{T}, \rt{B}, \rt{c} \}
        $$
    \item \textbf{Fully Connected} 
        Fully Connected layers share hardware with Convolution layers, but with no feature-map buffering. 
        The computation node is characterised as the following tuple:
        $$
            \Gamma = \{ \rt{\bm{S}}^{in}, \rt{\bm{S}}^{out}, \rt{c}^{in}, \rt{c}^{out} \}
        $$
\end{itemize}

\subsection{Hardware Design and Implementation} \label{sec:hw_design}

The proposed architecture follows the paradigm of a system consisting of a processor extended by a set of custom instructions. The core building blocks described before are equivalent to custom instructions, and their control is performed through a CPU. Each building block is connected to a crossbar that is responsible for handling the routing of data between the building blocks and off-chip memory, as well as performing inter building block routing. The memory access to and from memory is supported through dedicated DMA blocks.

The architecture of the custom instructions follows the Streaming Architecture paradigm, by implementing direct convolution operations and exploring the opportunity of driving the instantiated building blocks as dictated by the 3D model without accessing the off-chip memory, but at the same time can map multiple operations on the same building blocks in a time-shared manner avoiding the need for reconfiguring the FPGA fabric.

The toolflow is responsible to identify the required building blocks in order to support the computations of a given 3D CNN model, as well as to tune them in order to optimise the performance of the system given the available resources (FPGA resources and off-chip memory bandwidth). As such, the resulting system is a heterogeneous multi-core system, with blocks tailored to the 3D CNN model and the targeted FPGA device. This deviates from the approach taken in other streaming toolflows such as FINN \cite{Umuroglu2017}, and fpgaConvNet \cite{Venieris2019}, where the hardware is tailored to specific DNN models with layers being deeply pipelined, utilising bitstream reconfiguration to overcome resource constraints. Such design approach leads to high performance designs for throughput oriented applications, however bitstream reconfiguration inhibits the ability to target latency-driven applications with a latency target smaller than the reconfiguration time of the device, which is usually in the order of hundreds of milliseconds.

\begin{figure}[t]
    \centering
    \includegraphics[width=0.9\columnwidth]{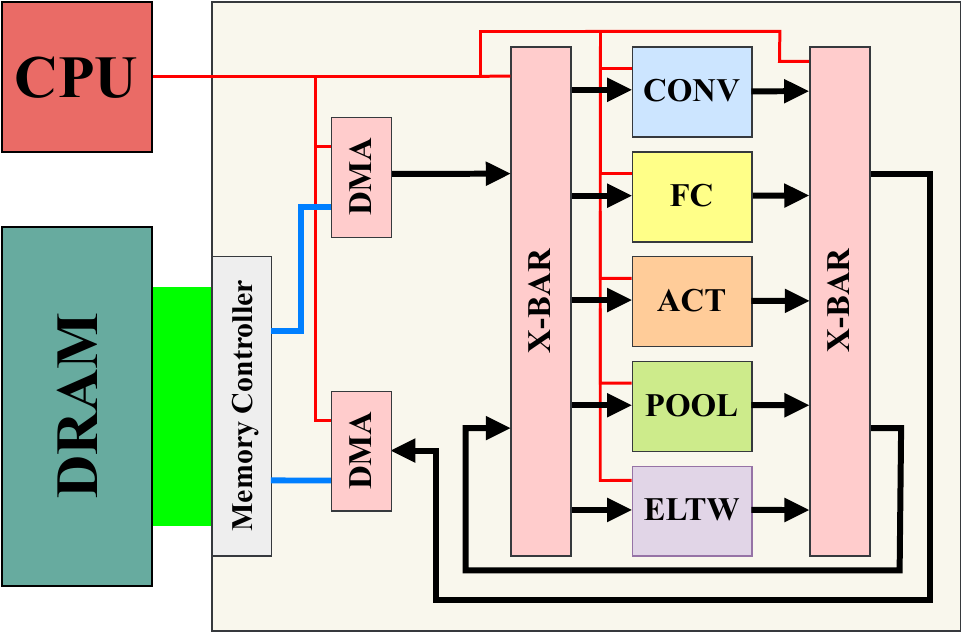}
    \caption{
        Block diagram of an accelerator instance produced by the HARFLOW3D toolflow. 
        The black lines describe AXI-Stream signals, where the arrows indicate the directionality of the connection,
        blue are high-throughput \textcolor{blue}{AXI} interfaces for DMA access,
        red are \textcolor{red}{AXI-Lite} connections for runtime configuration of the hardware nodes, and
        green indicate the \textcolor{green}{DDR IO} interfaces for communicating with off-chip memory.
    }

    \label{fig:system_diagram}
\end{figure}

Figure~\ref{fig:system_diagram} illustrates a simplified diagram of an example accelerator generated for a given model. 
Feature-maps are sent to and from off-chip memory via a pair of DMAs, and sent to the hardware nodes through the configurable crossbars. 
The design has a sandwich-like architecture, with the AXI-Stream crossbars routing data to and from hardware nodes. 
The output crossbar connects to the input crossbar, allowing for interconnectivity between hardware nodes.

\begin{figure}[t]
    \centering
    \includegraphics[width=\columnwidth]{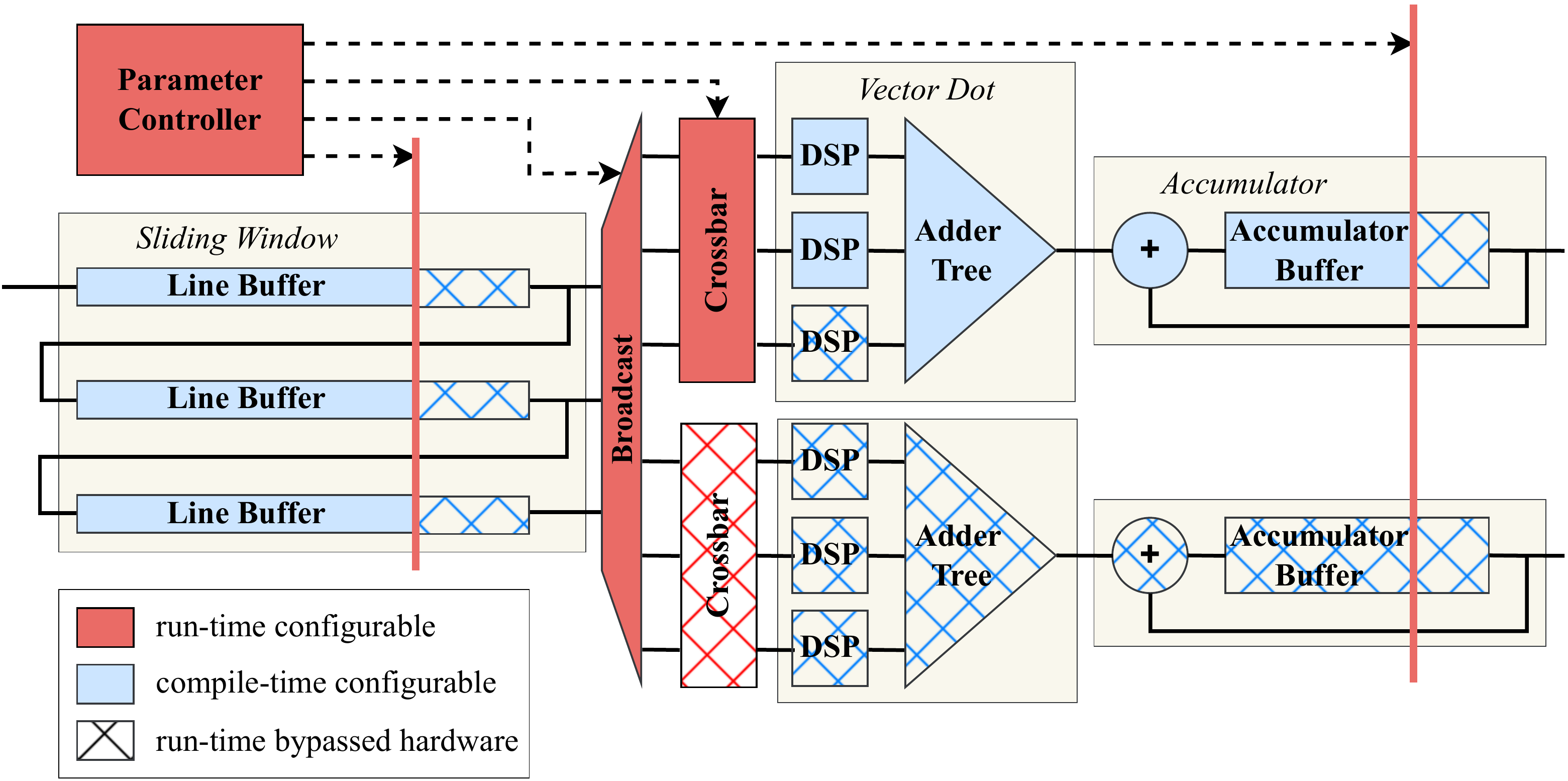}
    \caption{
        Diagram of hardware for Convolution, and how it can be used with runtime parameters. 
        The \textcolor{blue}{blue} blocks represent compile-time configurable hardware modules.
        The \textcolor{red}{red} blocks represent runtime configurable hardware modules.
        Cross-hatching gives an example of how hardware elements can be bypassed at runtime.
    }
    \label{fig:conv_diagram}
\end{figure}

The fpgaConvNet \cite{Venieris2019} toolflow is utilised for the implementation of the computation blocks. We regard this as the \textit{baseline} design, where no runtime configuration is supported, and only padded execution can be used to execute variable feature-map sizes.
Figure~\ref{fig:conv_diagram} illustrates how the baseline design has been modified to support runtime configurable layers. 
The highlighted blocks constitute the overhead required for supporting runtime reconfigurability. 
The additional hardware includes a configurable counter to change the depth of the line buffers and accumulation buffers, and a crossbar to map configurable kernel sizes to a configurable number of multipliers.
These extra resources are insignificant, and there is no change in targetable clock frequency.

%% file: Modelling/modelling.tex
\section{Modelling}
\label{sec:modelling}

Precise, high-quality performance and resource modelling is essential to support a rapid and valid design space exploration.
This section presents a comprehensive performance model for the parameterized building blocks described in Section III.
Performance models are described per computation node, and the resource model is for the entire system design.

\subsection{Performance Modelling}

Performance modelling is required to evaluate the latency objective during design space exploration.
The latency of the execution of a layer can be estimated by a roof-line model consisting of the required memory bandwidth, and the latency from computation. The latency model for each supported hardware type as a function of its runtime parameters $\Gamma$ is given as follows:

$$
	\mathcal{L}_{Conv}(\Gamma) = \frac{|\rt{\bm{S}}^{out}| \cdot \rt{F} \cdot |\rt{\bm{K}}|}{\rt{c}^{out} \cdot \rt{c}^{in} \cdot \rt{f}}
$$
$$
	\mathcal{L}_{FC}(\Gamma) = \frac{\rt{C} \cdot \rt{F} }{ \rt{c}^{in} \cdot \rt{c}^{out} }
$$
$$
	\mathcal{L}_{Pool}(\Gamma) = \mathcal{L}_{Act}(\Gamma) = \mathcal{L}_{EltWise}(\Gamma) = \frac{|\rt{\bm{S}}^{in}|}{\rt{c}}
$$

The preceding models assume unlimited memory bandwidth, however memory accesses have limited throughput.
In order to model the effect of memory bandwidth, the consumption and production rates of the layer are required, which are given as,
\begin{equation*}
	r_n^{in}(\Gamma) = \frac{ |\rt{\bm{S}}^{in}|  }{ \mathcal{L}_n(\Gamma) \cdot \rt{c}^{in}}, \;
	r_n^{out}(\Gamma) = \frac{ |\rt{\bm{S}}^{out}| }{ \mathcal{L}_n(\Gamma) \cdot \rt{c}^{out} }
\end{equation*}

where $r_n^{in}(\Gamma)$ and $r_n^{out}(\Gamma)$ are the words per cycle per stream in and out of the computation node $n$ when executing runtime parameters $\Gamma$.

For convolution and fully-connected layers in particular, extra memory bandwidth is required to stream in the weight parameters.
This rate is described as,

\begin{equation*}
	r_{Conv, FC}^{param}(\Gamma) = \frac{\rt{C}^{in} \cdot \rt{F} \cdot |\rt{\bm{K}}| }{ \mathcal{L}_{Conv, FC}(\Gamma) \cdot \rt{c}^{in} \cdot \rt{c}^{out} \cdot \rt{f}}
\end{equation*}

Alongside the double-buffering of weights, if the channel dimension of a convolution or full-connected layer is folded, the partial sums must be accumulated.
This requires streaming the previous partial sum from off-chip memory, whose rate matches the rate out.

\begin{equation*}
	r_{Conv, FC}^{psum}(\Gamma) = r_{Conv, FC}^{out}(\Gamma)
\end{equation*}

The constrained bandwidth in and out can be described as,

\begin{align*}
	\mathcal{B}_n^{in}(\Gamma)	&= \min \{ \mathcal{B}_{DMA}^{in}, \; r_n^{in}(\Gamma)	\cdot \rt{c}^{in} \} \\
	\mathcal{B}_n^{out}(\Gamma) &= \min \{ \mathcal{B}_{DMA}^{out},\; r_n^{out}(\Gamma) \cdot \rt{c}^{out} \}
\end{align*}

where $\mathcal{B}_n^{in}(\Gamma)$ and $\mathcal{B}_n^{out}(\Gamma)$ are the constrained words per cycle for the execution of parameters $\Gamma$ on the computation node $n$.
This describes a roofline model, where the on-chip bandwidth is capped by the memory bandwidth.
The above model summarises the bandwidth in and out for all layers apart from convolution and fully-connected.
To describe these, the bandwidth for parameters and partial sums must also be considered,

\begin{align*}
	\mathcal{B}^{in}_{Conv, FC}(\Gamma) = \min \{ \mathcal{B}_{DMA}^{in},& \;
	r^{in}(\Gamma) \cdot c^{in} + \\ &	r_{Conv, FC}^{psum}(\Gamma) \cdot c^{out} + \\ & r_{Conv, FC}^{param}(\Gamma) \cdot c^{in} \cdot c^{out} \cdot f \}
\end{align*}

For \textit{Fully Connected} layers, $f=1$.
Given the bandwidths, the total latency for executing a layer is given as,

\begin{equation}
\label{eq:layer_latency}
	\Tilde{\mathcal{L}}_n(\Gamma) = \max \left\{\frac{|\rt{\bm{S}}^{in}|}{\mathcal{B}_n^{in}(\Gamma)}, \; \frac{|\rt{\bm{S}}^{out}|}{\mathcal{B}_n^{out}(\Gamma)}\right\}
\end{equation}

It is worth noting that the performance models are functions of runtime parameters, owing to the highly customisable hardware.
Streaming Architectures tend to be computationally bounded, as supported by the results presented in Section ~\ref{sec:evaluation}. As the size of the feature-maps is significantly greater than that of the weights, the weights bandwidth is often negligible in comparison to the overall bandwidth required. There is no latency associated with updating the runtime parameters because they are also double-buffered and require negligible information transfer ($<$100B).

\subsection{Resource Modelling}
\label{sec:resource}

Resource modelling is used to explore the performance-resource design space whilst keeping designs within the target FPGA's constraints. Modern FPGA devices share four common resource types: DSP, BRAM, LUT and FF.
Only the \textit{Conv} and \textit{FC} layers use DSP resources,
An analytical model of DSP usage for building blocks $n$ of these layer types is given by,

\begin{equation*}
	\mathcal{R}_{Conv}^{DSP} = c_n^{in} \cdot c_n^{out} \cdot f_n, \; \; \mathcal{R}_{FC}^{DSP} = c_n^{in} \cdot c_n^{out} %
\end{equation*}
As a 16-bit fixed-point precision is used throughout the design, each DSP is used for either $16\cdot 16$ multiplication or multiplication-accumulation.
For BRAM modelling, the number of BRAM blocks can be described as,

\begin{equation*}
	\mathcal{R}^{BRAM}(depth, words) = \Bigl\lceil \frac{depth}{512} \Bigl\rceil \cdot \Bigl\lceil \frac{16 \cdot words}{36} \Bigl\rceil
\end{equation*}

The bus width of the required memory is $16 \cdot words$ as 16-bit fixed-point is used.
Only \textit{Conv}, \textit{FC} and \textit{Pool} layers consume BRAM components.
Both \textit{Conv} and \textit{Pool} use BRAM for the Sliding Window module, which is described as,

\begin{align*}
	\mathcal{R}_{SlW}^{BRAM} &= \mathcal{R}^{BRAM}\Bigl(W_n \cdot D_n \cdot \frac{C_n^{in}}{c_n^{in}}, (K_n^H-1) \cdot c_n^{in}\Bigl) \; + \\
								& \mathcal{R}^{BRAM}\Bigl( D_n \cdot \frac{C_n^{in}}{c_n^{in}}, K_n^H \cdot (K_n^W-1) \cdot c_n^{in}\Bigl) \; + \\
								& \mathcal{R}^{BRAM}\Bigl( \frac{C_n^{in}}{c_n^{in}}, K_n^H \cdot K_n^W \cdot (K_n^D-1) \cdot c_n^{in}\Bigl) \;
\end{align*}

\textit{Conv} and \textit{FC}, require some extra memory for storing weights on-chip, which is modelled as,

\begin{align*}
	\mathcal{R}_{Weight}^{BRAM} = \mathcal{R}^{BRAM}\Bigl( \frac{C_n^{in} \cdot F_n \cdot |\bm{K}_n| }{c_n^{in} \cdot c_n^{out} \cdot f_n}, c_n^{in} \cdot c_n^{out} \cdot f_n\Bigl)
\end{align*}

For \textit{Fully Connected} layers, $\bm{K}_n = \{1,1,1\}$ and $f_n=1$.
The hardware design uses a large data word design technique, which significantly improves BRAM utilisation.

For the modelling of LUT and FF resources, a regression model is used due to the non-deterministic nature of FPGA synthesis.
The regression models are obtained from a data set of 5000 synthesised modules, where the relationship between the module's parameters and resources are inferred.

Using the derived resource models, we can estimate the resource consumption of a complete hardware graph (G),

\begin{equation*}
	\bm{\mathcal{R}}_{total} = \left( \sum_{n \in G} \bm{\mathcal{R}}_{n} \right) + \bm{\mathcal{R}}_{DMA} + \bm{\mathcal{R}}_{xbar}
\end{equation*}

where $\bm{\mathcal{R}}$ describes the complete resources (DSP, BRAM, LUT, FF), for the given component.
The quality of the resource model is evaluated in Section~\ref{sec:validation}.

%% file: DesignSpaceExploration/dse.tex
\newcommand{\algcomment}[1]{%
	\vspace{0.05cm}%
	\noindent%
	{ #1\par}%
	\vspace{0.05cm}%
	}

\section{Latency-Driven Design Space Exploration}

The minimization of the 3D-CNN model's execution latency on the runtime-configurable accelerator is regarded as an optimization problem. As such, a Design Space Exploration is performed that searches for both an efficient accelerator for the specific application and a schedule for execution of the 3D CNN model layers on that architecture.

\subsection{Scheduling}

Once an accelerator design has been created, a schedule is needed for executing the 3D-CNN model's layers for this given design. The main choices with regard to scheduling are:

\begin{itemize}
	\item Mapping of the computation nodes for executing the 3D-CNN model's layers (i.e. execution nodes).
	\item Tiling of the feature-maps of execution nodes on a given computation node.
	\item Runtime configurations for all the invocations of the computation nodes based on the respective execution nodes parameters.
\end{itemize}

The mapping between a computation node and the respective execution nodes which it will execute is described as an execution mapping function $\mathcal{E} : G \mapsto \mathcal{P}(M)$, where $\mathcal{P}(M)$ is the power-set of $M$, which is all the distinct subsets of $M$.
This mapping creates disjoint subsets of $M$, which can be described as,

\begin{equation*}
	\mathcal{E}(n) \cap \mathcal{E}(m) = \emptyset \; \forall \; n, m \in G, n \neq m
\end{equation*}

where $\mathcal{E}$ is the execution mapping function.
The mapping function must give unique mappings for each computation node, such that none of the model's layers are executed more than once.
The inverse mapping, $\mathcal{E}^{-1}$ finds the corresponding computation node for a given execution node.
The mapping is decided based on the transform described in Section~\ref{sec:combine}.

Once a mapping has been decided, the schedule for $G$, denoted as $\Phi_G$, can be created for executing the 3D CNN model graph $M$ on the hardware graph $G$.
This schedule is outlined in Algorithm~\ref{alg:schedule}.
For each execution node $l$ of the 3D CNN model graph, the tiling factors across each of its dimensions are obtained, which are then used to find the tile sizes for computation.
The algorithm greedily allocates as much of the feature-map as possible onto the computation node and then chooses the coarse factors based on the tile shape.
Additional runtime parameters such as kernel sizes and padding are also chosen based on the execution node's parameters.

Having created the schedule $\Phi_G$, it can be used for executing the workload.
The ordering of dimensions in the proposed accelerator is $NHWDC$, where the channel dimension is the fastest changing, and so the schedule is also executed in this order.
The total latency for execution is described as,
\begin{equation}
	\mathcal{L}_{total}(G) = \sum_{n, \Gamma \in \Phi_G} \Tilde{\mathcal{L}}_n(\Gamma)
	\label{eq:perf_model}
\end{equation}

where $n$ is the computation node and $\Gamma$ is the corresponding set of parameters for each configuration in the schedule. The latency $\Tilde{\mathcal{L}}_n(\Gamma)$ of each node $n$ for a configuration $\Gamma$ is described in Equation~\eqref{eq:layer_latency}.

\begin{algorithm}[t]
\caption{Scheduling Algorithm}
\begin{algorithmic}[1]
\State $\Phi =$ empty list
\Comment{initialise an empty schedule}
\For{$l$ in $M$}
        \State $n = \mathcal{E}^{-1}(l)$
        \Comment{get the computation node}
        \For{$\bm{i}$ in range($\ceil{\frac{H^{in}_l}{H^{in}_n}},
								\ceil{\frac{W^{in}_l}{W^{in}_n}}, %
								\ceil{\frac{D^{in}_l}{D^{in}_n}}, %
								\ceil{\frac{C^{in}_l}{C^{in}_n}} )$} %

            \State $\begin{aligned}
			\rt{H} &= \min \{ H^{in}_n, H^{in}_l - i^{H} \cdot H^{in}_n \} \\
			\rt{W} &= \min \{ W^{in}_n, W^{in}_l - i^{W} \cdot W^{in}_n \} \\
			\rt{D} &= \min \{ D^{in}_n, D^{in}_l - i^{D} \cdot D^{in}_n \} \\
			\rt{C} &= \min \{ C^{in}_n, C^{in}_l - i^{C} \cdot C^{in}_n \} \\
            \end{aligned}$
            \If{ $type(n)$ \textbf{is} $Conv$ \textbf{or} $FC$ }
			\For{$\bm{i}^F$ in range($\ceil{\frac{F^{in}_l}{F^{in}_n}} )$}
			\State $\rt{F} = \min \{ F^{in}_n, F^{in}_l - i^{F} \cdot F^{in}_n \}$
			\State $c^{in} = \max \{ \; \textit{factors } \rt{C} \; \}$
			\State $c^{out} = \max \{ \; \textit{factors } \rt{F} \; \}$
			\State $\Gamma = \{ \rt{H}, \rt{W}, \rt{D}, \rt{C}, \rt{F}, \rt{c}^{in}, \rt{c}^{out} \}$
			\State $\Phi$ \textbf{append} ($n, \; \Gamma )$
			\EndFor
            \Else
			\State $c = \max \{ \textit{factors } \rt{C} \}$
			\State $\Gamma = \; \{ \rt{H}, \rt{W}, \rt{D}, \rt{C}, \rt{c} \}$
			\State $\Phi$ \textbf{append} ($n, \Gamma )$
            \EndIf
        \EndFor
\EndFor

\end{algorithmic}
\label{alg:schedule}
\end{algorithm}

\subsection{Optimization Strategy}

Simulated annealing (SA), a meta-heuristic for finding minima in non-convex functions, is adopted as the optimization approach for minimising the latency of a 3D CNN model to FPGA mapping.
The implementation of SA for this particular optimisation problem is given in Algorithm \ref{alg:annealing}.
\begin{algorithm}[b]
\caption{Simulated Annealing Optimisation Algorithm}
\begin{algorithmic}[1]
\State $\tau$ = $\tau_{start}$, $G_{new} = G_{init}$
\Comment{initialisation}
\While {$\tau \; > \; \tau_{min}$}
	\State $G_{prev}$ = $G_{new}$
	\Comment{store previous design}
	\State $\mathcal{L}_{prev} = \mathcal{L}_{total}(G_{prev})$
	\Comment{get design latency}
	\State $G_{new}$ = random transformations on $G_{prev}$
	\State $\mathcal{L}_{new} = \mathcal{L}_{total}(G_{new})$
	\Comment{get new design latency}
	\If{constraints satisfied}
		\If{ $\mathcal{L}_{new} < \mathcal{L}_{prev}$}
			\If{ $\psi (\mathcal{L}_{prev}, \mathcal{L}_{new}, \tau) < x \sim U(0,1)$}
				\State $G_{new}$ = $G_{prev}$
				\Comment{reject new design}
			\EndIf
		\EndIf
	\EndIf
	\State $\tau = \lambda \cdot \tau$
	\Comment{reduce temperature}
\EndWhile
\end{algorithmic}
\algcomment{where $\psi(\mathcal{L}_{prev}, \mathcal{L}_{new}, \tau) = \exp({-(\mathcal{L}_{prev} - \mathcal{L}_{new})/\tau})$}
\label{alg:annealing}
\end{algorithm}

The main objective of the algorithm is to seek the global minimum of a given cost function; in this case, the cost function is latency, which is derived from Equation~\eqref{eq:perf_model}.
Starting at an initial state by producing random values for the parameters of the computation nodes, transformations to the hardware graph $G$ are applied iteratively generating new states that may be accepted or rejected based on a policy described in Algorithm \ref{alg:annealing}.
Each new state is evaluated based on the system's predetermined constraints, and is only accepted if it satisfies all of them.
The constraints that must be satisfied on the proposed system are the following:
\begin{itemize}
	\item The available memory bandwidth should not be exceeded.
	\item The total used resources $\bm{\mathcal{R}}_{total}$ should not exceed the available resources of the device.
	\item The streams in and out ($c_n^{in}, c_n^{out}$) of each computation node should be a factor of the channels in and out respectively.
	\item The derived scheduled parameters of each layer must be less than the maximum supported by the computation node.
\end{itemize}

\begin{figure}
    \centering
    \includegraphics[width=\columnwidth, keepaspectratio]{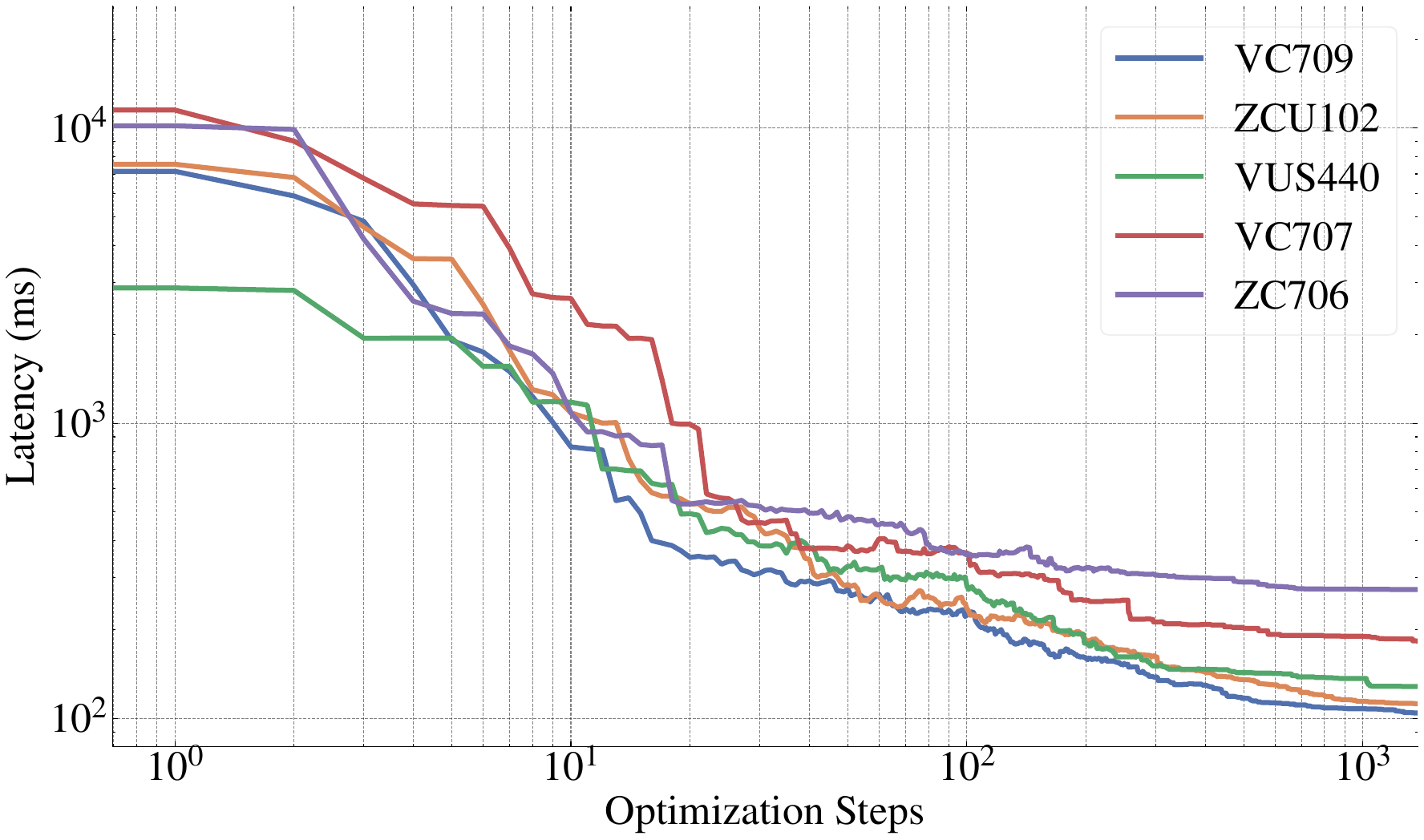}
    \caption{
        Evolution of latency during Simulated Annealing for various FPGA devices. 
    }
    \label{fig:latency_distribution}
    \vspace{-0.3cm}
\end{figure}

Figure \ref{fig:latency_distribution} depicts the evolution of latency during Simulated Annealing.
The graph depicts how the latency of the C3D model on various FPGA devices evolves over time.
As indicated by the graph, the starting point for each run has extremely high latency, as all of the parameters are set to random values.
The latency continues to improve until it reaches a plateau, at which point the optimization is terminated.

\subsection{Transformations}
\label{sec:transforms}

In order to traverse the design space, a set of transforms on the hardware graph $G$ are described below.

\subsubsection{Feature-Map Dimensions Reshaping} %

The feature-map dimensions dictate the amount of on-chip memory required both for the weight parameters as well as caching required by the Sliding Window module, as given in Section~\ref{sec:resource}.
At compile time, a fixed shape for the feature-map dimensions must be given to the computation node, and at run time the layer can be executed on the computation node by tiling the feature-map.

When the optimizer searches for the optimal feature-map shape configuration for the computation node for both the input and output, the following conditions must hold,
\begin{align*}
	D_{n} &\leq max \{ D_{l} : l \in M \} \\
	H_{n} &=	max \{ H_{l} : l \in M \} \\
	W_{n} &\leq max \{ W_{l} : l \in M \} \\
	C_{n} &\in \{ \textit{ factors } C_{l} : l \in M \}
\end{align*}

As the choice of row dimension has no impact on resources, the maximum of all rows is chosen.
For depth and columns, any dimension that is both less than the max of all layers, and greater than a minimum feasible dimension is acceptable.
The channel dimension is chosen to be a factor of any of the existing channel dimensions.

\subsubsection{Coarse-grain Folding}
The coarse-grain folding transformation modifies the number of parallel executions of coarse operations in each layer to achieve parallelism over the channel dimensions of the input feature-map. The primary operations of each layer can be performed simultaneously by deploying (at max) as many instances of its processing blocks as the number of channels. On Fully Connected and 3D Convolutional layers the coarse-level parallelism can be utilised at both the input and output channels.
This parallelism is achieved by updating and searching for appropriate values of the compile-time parameters $c_n^{in}$, $c_n$, and $c_n^{out}$ during optimization, which are also taken into account by the performance and resource models. To be considered valid, a design must comply with the constraints driving this transformation.
~
\begin{equation*}
	c_n^{in},c_n \in \textit{factors } C_n^{in}, \; c_n^{out} \in \textit{factors } C_n^{out}
\end{equation*}

\subsubsection{Fine-grain Folding}
The second folding-wise transformation factor determines the parallelism of the vector dot product operation for 3D convolutional layers. This sort of parallelism specifies the number of multipliers to be set in parallel for multiplications and the number of levels on the adder trees for additions. By increasing the compile-time fine folding factor, achievable latency is reduced at the cost of extra DSP resources, as described in Section~\ref{sec:modelling}. Evidently, there is a trade-off between performance and resource utilization. This type of parallelism is accomplished by modifying the $f_n$ parameter during the optimization process. The constraint on the compile-time fine-grain folding $f_n$ is given as,
~
\begin{equation*}
	f_n \in \textit{factors } |\bm{K}_n|
\end{equation*}

\subsubsection{Combination and Separation of Computation Nodes}
\label{sec:combine}
As stated in Section \ref{sec:hw_design}, the toolflow allows several model execution nodes to share the same computation node.
The initial mapping creates unique computation nodes $n$ for each execution node $l$.
In NNs with multiple layers, this is impractical since the FPGA resources can be quickly exhausted, and the performance of each computation node would be compromised in order to fit the design. A combination of execution nodes by type for a single computation node (the available types are depicted in Figure~\ref{fig:system_diagram}) is proposed as a solution to this issue. All execution nodes of the same type are combined and mapped onto a single computation node at the beginning of the optimization. The compile-time parameters of this node are then modified such that it can handle the workload of its associated execution nodes. This transformation is employed throughout the optimization procedure, and can affect the computation and execution nodes in two ways:
\begin{itemize}
	\item \underline{Separate Computation Nodes:} The algorithm chooses $L_e$ execution nodes, where $L_e$ is a hyperparameter, and detaches them from their corresponding computation node. The new group of execution nodes are integrated and mapped onto new computation nodes whose characteristics and parameters are adapted respectively.
	\item \underline{Combine Computation Nodes:} The algorithm searches for computation nodes of the same type and selects $N_c$ of them, where $N_c$ is a hyperparameter, to combine into a single computation node.
	The computation node's compile-time parameters are updated to support the new set of workloads.
\end{itemize}
Each time the combination or separation is applied, a set of constraints are considered to assure the result's validity. The required constraints are a combination of the \textit{Feature-Map Dimensions Reshaping, Coarse-grain Folding, and Fine-grain Folding} constraints.

\begin{figure}
	\centering
	\begin{subfigure}[b]{\columnwidth}
		\centering
		\includegraphics[scale=0.5]{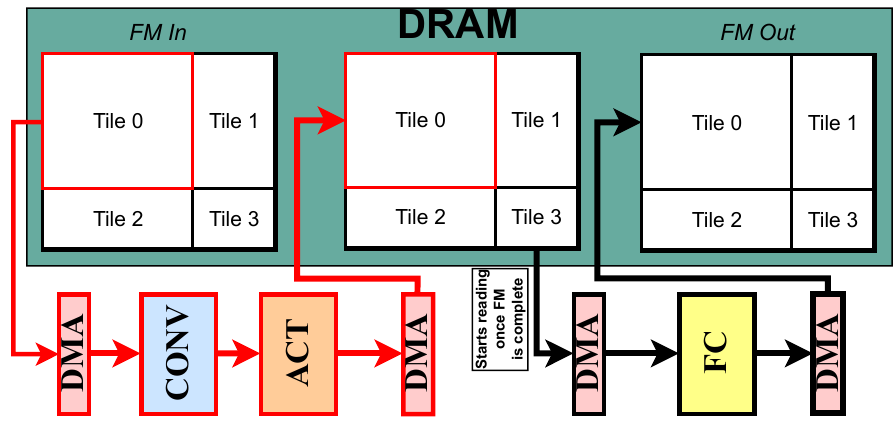}
		\caption{3D CNN Model to SDFG mapping}
		\label{fig:hardware_graph_horizontal}
	\end{subfigure}
	\begin{subfigure}[b]{0.4\textwidth}
		\centering
		\includegraphics[scale=0.5]{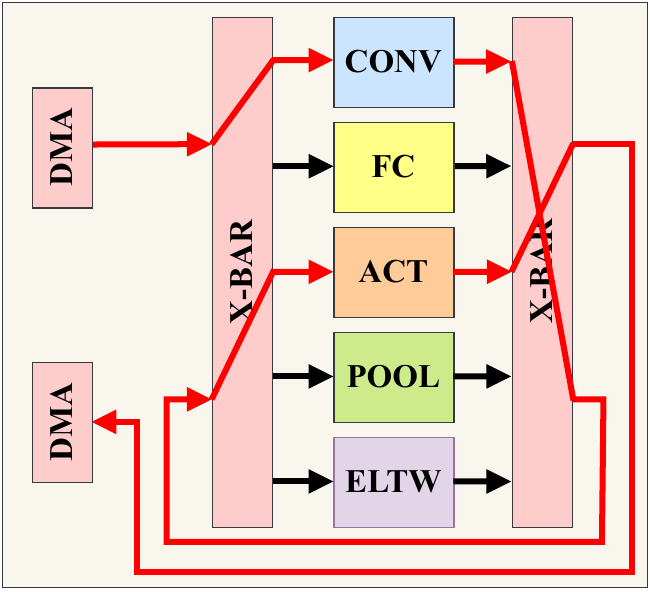}
		\caption{Hardware execution}
		\label{fig:hardware_mapping}
	\end{subfigure}
	\caption{The Dataflow of a simple design consisting of Convolution, ReLU, and FC layers. As the red lines and the crossbar dictate the flow between Convolution and ReLU can be addressed within the FPGA without sending the data back to the off-chip memory. This is the result of Fuse Activation optimization.}
	\label{fig:model_mapping}
\end{figure}

%% file: Validation/validation.tex
\section{Hardware Model Validation}
\label{sec:validation}

Modelling of performance and resources is used extensively for rapid traversal of the large design space.
In this section, the accuracy of the performance and resource models is validated across different hardware designs for C3D~\cite{Ji2013}, demonstrating negligible error between modelled and measured results.

\begin{table*}
\caption{
	Comparison of predicted and synthesised resources for C3D designs on a ZCU102 board.
	}
\label{tab:rsc_validation}
\centering
\begin{tabular}{@{}crrrrrrrrrrrrr@{}}
\toprule
	\multirow{2}{*}{\textbf{\begin{tabular}[c]{@{}c@{}}Hardware\\ Node\end{tabular}}} &
	\multicolumn{3}{c}{\textbf{DSP}} &
	\multicolumn{3}{c}{\textbf{BRAM}} &
	\multicolumn{3}{c}{\textbf{LUT}} &
	\multicolumn{3}{c}{\textbf{FF}} \\

	\textit{} &
	\textit{pred.} &
	\textit{act.} &
	\textit{error} &
	\textit{pred.} &
	\textit{act.} &
	\textit{error} &
	\textit{pred.} &
	\textit{act.} &
	\textit{error} &
	\textit{pred.} &
	\textit{act.} &
	\textit{error}
   \\ \midrule \midrule
	Conv &
		2304 & 2304 & {\scriptsize(+0\%)}	 &
		1052 & 1052 & {\scriptsize(+0\%)} &
		151K & 138K & {\scriptsize(+9.4\%)}  &
		155K & 166K & {\scriptsize(-6.6\%)}  \\
	MaxPool &
		0	 & 0	& {\scriptsize(+0\%)} &
		0	 & 0	& {\scriptsize(+0\%)}	 &
		22K  & 17K	& {\scriptsize(+29.4\%)} &
		16K  & 18K	& {\scriptsize(-11.1\%)} \\
	Gemm &
		128  & 128	& {\scriptsize(+0\%)} &
		456  & 456	& {\scriptsize(+0\%)}  &
		11K  & 10K	& {\scriptsize(+1.0\%)} &
		15K  & 18K	& {\scriptsize(-16.6\%)} \\
  ReLU &
		0	 & 0	& {\scriptsize(+0\%)}  &
		0	 & 0	& {\scriptsize(+0\%)}  &
		1.0K & 1.4K & {\scriptsize(-28.5)} &
		2.2K & 2.2K & {\scriptsize(+0\%)}  \\
\midrule
	DMA &
		  &  0 &   &
		  & 51 &   &
		  & 2.9K &	 &
		  & 4.7K &	 \\
	X-BAR &
		  & 0 &   &
		  & 0 &   &
		  & 1.7K &	 &
		  & 1.4K &	 \\
\midrule
\textbf{Total} &
		2432 & 2432 & {\scriptsize(+0\%)} &
		1559 & 1559 & {\scriptsize(+0\%)} &
		189K & 171K & {\scriptsize(+7.8\%)}  &
		194K & 210K & {\scriptsize(-9.4\%)}  \\
\textit{\textbf{Avail.}} &
		\multicolumn{3}{c}{\textit{(2520)}} &
		\multicolumn{3}{c}{\textit{(1824)}} &
		\multicolumn{3}{c}{\textit{(274K)}} &
		\multicolumn{3}{c}{\textit{(548K)}} \\
\bottomrule
\end{tabular}
\end{table*}

Table~\ref{tab:rsc_validation} shows a direct comparison between predicted resources and resources after synthesis for a C3D design.
The DSP and BRAM models are highly accurate due to the deterministic nature of their synthesis, as resource type annotations are used in the hardware design.
For the LUT and FF prediction accuracy, the modelling over-predicts LUT usage and under-predicts FF usage.
Logic optimisation contributes to fewer LUTs in the final implemented design, and the additional FF resources likely arise from inter-module buffering that is neglected in the modelling. %
Additional BRAM resources are required by DMAs for the buffering of bursts across the feature-map. This is accounted for during optimisation.

The convolution layers dominate the total resource consumption, with DSPs typically being the limiting factor.
The convolution layers are investigated further, where statistical resource modelling information is captured over 16 designs with varying configurations among different layers.
Supporting the results of Table \ref{tab:rsc_validation}, the Mean Absolute Percentage Error (MAPE) and the Standard Deviation ($\sigma$) over the 16 different convolution configurations are shown in Table \ref{tab:rsc_statistics}.

\begin{table}[H]
\centering
\caption{Statistical resource modelling information over multiple runs for different convolution layers and configurations.}
\label{tab:rsc_statistics}
\resizebox{0.8\columnwidth}{!}{%
\begin{tabular}{lllll}
\hline
		  & \multicolumn{1}{l|}{DSP} & \multicolumn{1}{l|}{BRAM} & \multicolumn{1}{l|}{LUT}  & FF	\\ \hline \hline
MAPE (\%) & \multicolumn{1}{l|}{0.0} & \multicolumn{1}{l|}{0.35} & \multicolumn{1}{l|}{7.21} & 8.81 \\
$\sigma$	   & \multicolumn{1}{l|}{0.0} & \multicolumn{1}{l|}{0.38} & \multicolumn{1}{l|}{8.82} & 2.89 \\ \hline
\end{tabular}%
}
\end{table}

\begin{figure}
    \centering
    \includegraphics[width=\columnwidth, keepaspectratio]{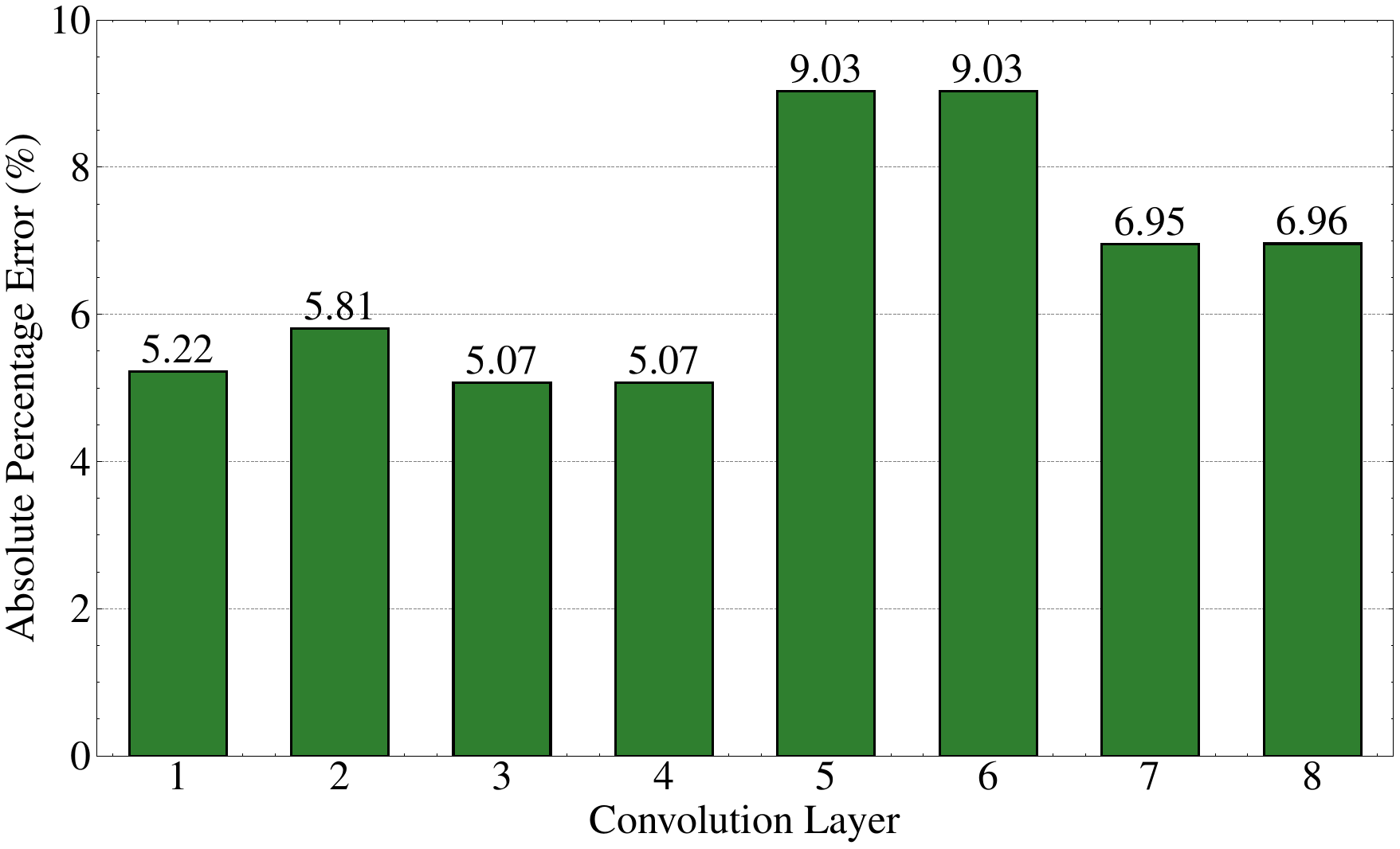}
    \caption{
        Comparison of predicted latency to measured latency for all the convolution layers of C3D on the ZCU106 board, depicted as absolute percentage error.
    }
    \label{fig:percentage_error_zcu106}
\end{figure}

The accuracy of the performance model is evaluated in Figure~\ref{fig:percentage_error_zcu106}, where the calculation of the absolute percentage error is given by: $error = \frac{\lvert Predicted-Measured \rvert}{Measured}*100$.
The small percentage differences between predicted and measured results on ZCU106 board imply a good level of accuracy and confidence in the toolflow's modelling results.
The divergence between the expected and actual latency of the layers is due to the DMA introducing a delay between bursts due to memory access cycles.
The MAPE for all convolution layers of the C3D model at the particular design is 6.64\%.

%% file: Evaluation/evaluation.tex
\section{Evaluation}
\label{sec:evaluation}

This section focuses on the evaluation of the proposed methodology and its ability to discover optimal designs.
The automatically generated designs are benchmarked against existing 3D CNN accelerators.

The models of 3D CNN included in the evaluation are listed in Table~\ref{tab:models_characteristics}.
Each of these models has demonstrated state-of-the-art performance in a number of HAR benchmarks and offers a variety of workload and network parameters. Some of these models also serve as benchmarks for existing FPGA accelerator-focused works. The ONNX files used for all the experimental results have been exported from \cite{MMAction22020OpenMMLabsBenchmark} for \textit{C3D, Slowonly}, and \textit{X3D-M} models, and from \cite{Hara2018CanImageNet} for \textit{R(2+1)D-18} and \textit{R(2+1)D-34}.

\begin{table}
\centering
\caption{Characteristics of the evaluated 3D CNN models.}\label{tab:models_characteristics}
\resizebox{\columnwidth}{!}{%
\begin{threeparttable}
\begin{tabular}{cc:c:c:c:c}
\hline
								& C3D		& Slowonly		& R(2+1)D-18		& R(2+1)D-34		& X3D-M \\
\hline
FLOPs (G)\tnote{$\dagger$}						 & 38.61	 & 54.81		 & 8.52				 & 12.91			 & 6.97  \\
Parameters (M)					& 78.41		& 32.51			& 33.41				& 63.72				& 3.82	\\
Num. of Layers					& 27		& 174			& 82				& 154				& 396	\\
Num. of Conv Layers				& 8			& 53			& 37				& 69				& 115	\\
Spatial dimensions				& $112 \times 112$	 & $256 \times 256$    & $112 \times 112$	 & $112 \times 112$    & $256 \times 256$	 \\
Num. of Frames					& 16		& 8				& 16				& 16				& 16	\\
\begin{tabular}[c]{@{}l@{}}UCF101\\ Accuracy (\%)\end{tabular}			& 83.2		& 94.54			& 88.66				& 92.27				& 96.52 \\
\hline
\end{tabular}
\begin{tablenotes}
	\item[$\dagger$] FLOPs are reported as MAC operations.
\end{tablenotes}
\end{threeparttable}
}
\end{table}

\subsection{Experimental Results}

\begin{table*}
\caption{Comparison of HARFLOW3D with existing works, for 3D CNN HAR models. Our tool can target all models, including X3D.}
\label{tab:performance_comparison}
\resizebox{\textwidth}{!}{%
\begin{threeparttable}
\begin{tabular}{@{}ccccccccccccccccccccc@{}}
\toprule

& H. Fan{\cite{Fan2017}} & H. Fan{\cite{Fan2018}} & Z. Liu{\cite{Liu2019AFpgas}} & T. Teng{\cite{Teng2020ExplorationAccelerator}} & \multicolumn{2}{c}{J. Shen \cite{Shen2018}\tnote{$\ddagger$}} & \multicolumn{2}{c}{M. Sun\cite{Sun20203DPruning}} & H. Fan{\cite{Fan2019}} & F. H. Khan{\cite{Khan2023TowardsNetworks}} & \multicolumn{10}{c}{HARFLOW3D} \\

Architecture & Hand-Tuned & Hand-Tuned & Partial\tnote{$\ast$} & Hand-Tuned & \multicolumn{2}{c}{Partial\tnote{$\ast$}} & \multicolumn{2}{c}{Partial\tnote{$\ast$}} & Hand-Tuned & Hand-Tuned & \multicolumn{10}{c}{Toolflow} \\ \midrule \midrule

Model & \multicolumn{1}{c|}{C3D} & \multicolumn{1}{c|}{C3D} & \multicolumn{1}{c|}{C3D} & \multicolumn{1}{c|}{C3D} & \multicolumn{2}{c|}{C3D} & C3D & \multicolumn{1}{c|}{R(2+1)D-18} & \multicolumn{1}{c|}{E3D} & \multicolumn{1}{c||}{I3D} & \multicolumn{2}{c|}{C3D} & \multicolumn{2}{c|}{Slowonly} & \multicolumn{2}{c|}{R(2+1)D-18} & \multicolumn{2}{c|}{R(2+1)D-34} & \multicolumn{2}{c}{X3D-M} \\

GFLOPs\tnote{$\dagger$} & \multicolumn{1}{c|}{38.61} & \multicolumn{1}{c|}{38.61} & \multicolumn{1}{c|}{38.61} & \multicolumn{1}{c|}{38.61} & \multicolumn{2}{c|}{-} & 38.61 & \multicolumn{1}{c|}{8.52} & \multicolumn{1}{c|}{6.1} & \multicolumn{1}{c||}{110} & \multicolumn{2}{c|}{38.61} & \multicolumn{2}{c|}{54.81} & \multicolumn{2}{c|}{8.52} & \multicolumn{2}{c|}{12.91} & \multicolumn{2}{c}{6.97} \\

\rowcolor{lightgray} Accuracy (\%) & \multicolumn{1}{c|}{79.87} & \multicolumn{1}{c|}{81.99} & \multicolumn{1}{c|}{83.2} & \multicolumn{1}{c|}{83.2} & \multicolumn{2}{c|}{83.2} & 83.2 & \multicolumn{1}{c|}{88.66} & \multicolumn{1}{c|}{85.17} & \multicolumn{1}{c||}{95} & \multicolumn{2}{c|}{83.2} & \multicolumn{2}{c|}{94.54} & \multicolumn{2}{c|}{88.66} & \multicolumn{2}{c|}{92.27} & \multicolumn{2}{c}{96.52} \\

FPGA & \multicolumn{1}{c|}{ZC706} & \multicolumn{1}{c|}{ZC706} & \multicolumn{1}{c|}{VC709} & \multicolumn{1}{c|}{VC707} & VC709 & \multicolumn{1}{c|}{VUS440} & ZCU102 & \multicolumn{1}{c|}{ZCU102} & \multicolumn{1}{c|}{Intel SX660} & \multicolumn{1}{c||}{VC709} & ZCU102 & \multicolumn{1}{c|}{VC709} & ZCU102 & \multicolumn{1}{c|}{VC709} & ZCU102 & \multicolumn{1}{c|}{VC709} & ZCU102 & \multicolumn{1}{c|}{VC709} & ZCU102 & \multicolumn{1}{c}{VC709} \\

\rowcolor{lightgray} Latency/clip & \multicolumn{1}{c|}{542.5} & \multicolumn{1}{c|}{476.8} & \multicolumn{1}{c|}{115.5} & \multicolumn{1}{c|}{107.9} & 89.4 & \multicolumn{1}{c|}{49.1} & 487 & \multicolumn{1}{c|}{243} & \multicolumn{1}{c|}{35.32} & \multicolumn{1}{c||}{96} & 98.15 & \multicolumn{1}{c|}{91.03} & 309.56 & \multicolumn{1}{c|}{239.34} & 48.99 & \multicolumn{1}{c|}{46.02} & 70.05 & \multicolumn{1}{c|}{62.55} & 155.07 & \multicolumn{1}{c}{120.38} \\

GOps/s & \multicolumn{1}{c|}{71.17} & \multicolumn{1}{c|}{80.97} & \multicolumn{1}{c|}{334.28} & \multicolumn{1}{c|}{357.83} & 431.87 & \multicolumn{1}{c|}{786.35} & 79.28 & \multicolumn{1}{c|}{35.06} & \multicolumn{1}{c|}{172.8} & \multicolumn{1}{c||}{1145.83} & 393.37 & \multicolumn{1}{c|}{424.14} & 177.05 & \multicolumn{1}{c|}{229.01} & 173.91 & \multicolumn{1}{c|}{185.13} & 184.29 & \multicolumn{1}{c|}{206.39} & 43.78 & 56.14 \\

GOps/s/DSP & \multicolumn{1}{c|}{0.079} & \multicolumn{1}{c|}{0.089} & \multicolumn{1}{c|}{0.092} & \multicolumn{1}{c|}{0.127} & 0.119 & \multicolumn{1}{c|}{0.273} & 0.031 & \multicolumn{1}{c|}{0.013} & \multicolumn{1}{c|}{0.102} & \multicolumn{1}{c||}{0.318} & 0.156 & \multicolumn{1}{c|}{0.117} & 0.07 & \multicolumn{1}{c|}{0.063} & 0.069 & \multicolumn{1}{c|}{0.051} & 0.073 & \multicolumn{1}{c|}{0.057} & 0.017 & 0.015 \\

Op/DSP/cycle & \multicolumn{1}{c|}{0.459} & \multicolumn{1}{c|}{0.449} & \multicolumn{1}{c|}{0.773} & \multicolumn{1}{c|}{0.798} & 0.799 & \multicolumn{1}{c|}{1.365} & 0.209 & \multicolumn{1}{c|}{0.092} & \multicolumn{1}{c|}{0.68} & \multicolumn{1}{c||}{1.59} & 0.781 & \multicolumn{1}{c|}{0.785} & 0.351 & \multicolumn{1}{c|}{0.424} & 0.345 & \multicolumn{1}{c|}{0.342} & 0.365 & \multicolumn{1}{c|}{0.382} & 0.086 & 0.104 \\

Frequency (MHz) & \multicolumn{1}{c|}{172} & \multicolumn{1}{c|}{200} & \multicolumn{1}{c|}{120} & \multicolumn{1}{c|}{160} & 150 & \multicolumn{1}{c|}{200} & 150 & \multicolumn{1}{c|}{150} & \multicolumn{1}{c|}{150} & \multicolumn{1}{c||}{200} & 200 & \multicolumn{1}{c|}{150} & 200 & \multicolumn{1}{c|}{150} & 200 & \multicolumn{1}{c|}{150} & 200 & \multicolumn{1}{c|}{150} & 200 & 150 \\

Precision & \multicolumn{1}{c|}{\begin{tabular}[c]{@{}c@{}}fp-16\end{tabular}} & \multicolumn{1}{c|}{BFP} & \multicolumn{1}{c|}{\begin{tabular}[c]{@{}c@{}}fp-16\end{tabular}} & \multicolumn{1}{c|}{\begin{tabular}[c]{@{}c@{}}fp-8\end{tabular}} & \begin{tabular}[c]{@{}c@{}}fp-16\end{tabular} & \multicolumn{1}{c|}{\begin{tabular}[c]{@{}c@{}}fp-16\end{tabular}} & \begin{tabular}[c]{@{}c@{}}fp-16\end{tabular} & \multicolumn{1}{c|}{\begin{tabular}[c]{@{}c@{}}fp-16\end{tabular}} & \multicolumn{1}{c|}{\begin{tabular}[c]{@{}c@{}}float-32\end{tabular}} &
\multicolumn{1}{c||}{\begin{tabular}[c]{@{}c@{}}fp-8\end{tabular}} & \begin{tabular}[c]{@{}c@{}}fp-16\end{tabular} & \multicolumn{1}{c|}{\begin{tabular}[c]{@{}c@{}}fp-16\end{tabular}} & \begin{tabular}[c]{@{}c@{}}fp-16\end{tabular} & \multicolumn{1}{c|}{\begin{tabular}[c]{@{}c@{}}fp-16\end{tabular}} & \begin{tabular}[c]{@{}c@{}}fp-16\end{tabular} & \multicolumn{1}{c|}{\begin{tabular}[c]{@{}c@{}}fp-16\end{tabular}} & \begin{tabular}[c]{@{}c@{}}fp-16\end{tabular} & \multicolumn{1}{c|}{\begin{tabular}[c]{@{}c@{}}fp-16\end{tabular}} & \begin{tabular}[c]{@{}c@{}}fp-16\end{tabular} & \begin{tabular}[c]{@{}c@{}}fp-16\end{tabular}\\

DSP (\%) & \multicolumn{1}{c|}{90} & \multicolumn{1}{c|}{86.6} & \multicolumn{1}{c|}{99.8} & \multicolumn{1}{c|}{96} & 42 & \multicolumn{1}{c|}{53} & 48 & \multicolumn{1}{c|}{48} & \multicolumn{1}{c|}{93.3} & \multicolumn{1}{c||}{100} & 96.51 & \multicolumn{1}{c|}{97.77} & 62.38 & \multicolumn{1}{c|}{69.11} & 96.94 & \multicolumn{1}{c|}{97.91} & 95.67 & \multicolumn{1}{c|}{97.25} & 53.52 & 89.61\\

BRAM (\%) & \multicolumn{1}{c|}{86.6} & \multicolumn{1}{c|}{88.1} & \multicolumn{1}{c|}{26.6} & \multicolumn{1}{c|}{25.3} & 52 & \multicolumn{1}{c|}{30} & 100 & \multicolumn{1}{c|}{100} & \multicolumn{1}{c|}{-} & \multicolumn{1}{c||}{79} & 71.93 & \multicolumn{1}{c|}{63.33} & 78.56 & \multicolumn{1}{c|}{81.05} & 69.13 & \multicolumn{1}{c|}{64.62} & 69.24 & \multicolumn{1}{c|}{54.52} & 56.14 & 78.67 \\

\bottomrule
\end{tabular}%
\begin{tablenotes}
	\item[$\ast$] Proposed design supports multiple models (both 2D and 3D), although being tailored to the characteristics of specific 3D CNN models.
	\item[$\dagger$] FLOPs are reported as MAC operations.
        \item[$\ddagger$] The C3D model used is different/smaller version from the original one \cite{Ji2013}.
\end{tablenotes}
\end{threeparttable}
}
\end{table*}

\subsubsection{Ablation Study}

In this section, an ablation study was undertaken to assess the effect of several optimization strategies on the final performance of the proposed method. Having a baseline strategy in place, as well as introducing and investigating modifications and additions to it, has yielded significant insights and conclusions regarding the direction that should be followed for improving the optimization process. For the ablation experiments, the R(2+1)D-18 model was used, although the findings apply to all supported models and devices.

\textbf{Baseline Design Description:}\\
The baseline optimisation strategy is defined as follows: the SA hyperparameters are configured by, $\tau_{start}=10$, $\tau_{min}=\num{1e-06}$, and the cooling rate $\lambda=0.99$, while a warm start is executed prior to the execution of the optimiser. These parameters remain constant throughout the ablation study. For the baseline experiments, the \textit{Feature-Map Dimensions Reshaping, Coarse-grain Folding, and Fine-grain Folding} transformations are enabled while the use of runtime parameters and the fusion of activation layers to preceding layer are disabled.

\textbf{Optimization Strategies:}
\begin{itemize}
	\item \underline{Building Blocks Combination}: Incorporating the \textit{Combination and Separation of Computation Nodes} transformation into the optimization approach improved the optimiser's performance by $1.14\times$. This optimization is described in detail in Section \ref{sec:transforms}.
	\item \underline{Fusion of activation functions into previous layer}: Upon revisiting the optimization findings and assessing each layer type of the model, it was found that the activation layers are typically memory bounded, which hinders the overall performance of the design.
	To overcome this constraint, the fusion optimization was introduced.
	Through the crossbars seen in Figure \ref{fig:hardware_mapping}, such layers are fused directly to the preceding ones (mostly convolution layers).
	Since convolution layers are mainly compute bounded, the fused activation layers also become compute bounded, suppressing the memory bounded limitation.
	The addition of this specific optimization has provided a \textbf{$1.52\times$} boost in performance.
	\item \underline{Runtime reconfiguration of layer parameters}: The introduction of runtime reconfigurability of computation nodes resulted in the most significant improvement boost. As described in Section \ref{sec:hw_design}, in a non-runtime parameterizable node, in order to support a layer with different runtime feature-map dimensions the underlying hardware modules would need to add padding to match the compile-time dimensions. This affects the performance of the computation node greatly, as it must perform redundant operations. With the introduction of runtime parameterisable modules, as shown in Figure \ref{fig:conv_diagram}, the extra cost of padding and all the redundant operations are omitted, resulting in a \textbf{$18.21\times$} performance increase.
\end{itemize}

\subsubsection{Resources against Latency Comparison}

\begin{figure}
	\centering
	\includegraphics[width=\columnwidth,keepaspectratio]{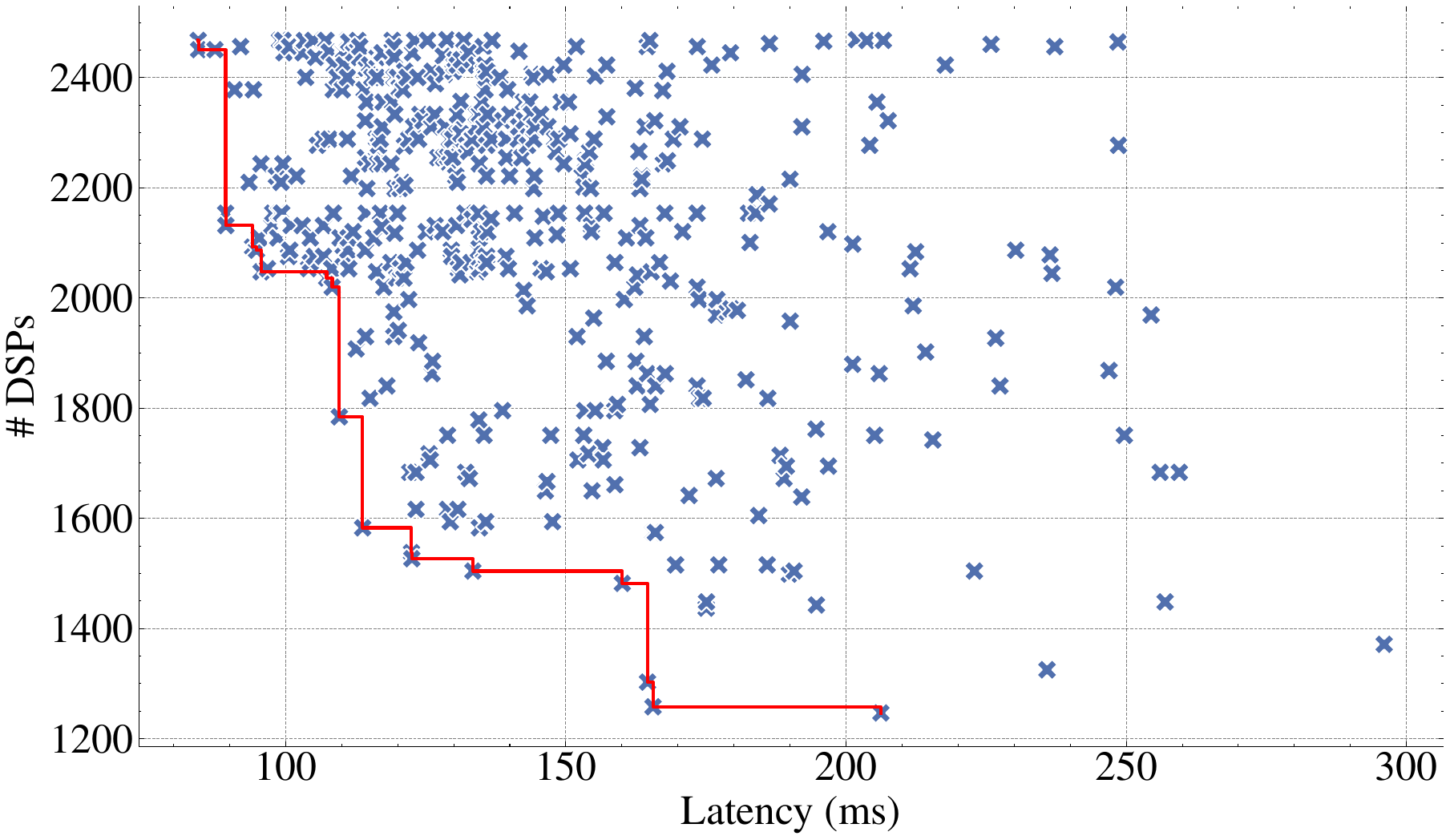}
	\caption{Pareto front of DSP utilization against latency for the R(2+1)D-34 model on a ZCU102 platform during Simulated Annealing optimization.}
	\label{fig:rsc_latency_pareto}
\end{figure}

Figure~\ref{fig:rsc_latency_pareto} depicts the pareto-front between resources and latency.
The DSP utilisation was chosen to represent resources, as the generated designs are typically limited by the number of available DSPs. %
The figure shows the optimiser's ability to traverse the resource-latency design space, achieving fine-grain control over this trade-off.
Indeed the optimiser is able to double performance along the pareto front at the cost of double the number of DSPs, exploring many design points in between.

\subsection{Comparison to the state-of-the-art}

Having evaluated the performance of the hardware and the effectiveness of the design space exploration method, the proposed work is positioned against existing accelerator works.
Table~\ref{tab:performance_comparison} outlines the current space for FPGA-based 3D CNN acceleration, and how HARFLOW3D compares.
Please note that all existing accelerators are the product of approaches that target only the specific workload, and thus are hand-tuned, where the produced designs from HARDFLOW3D are generated by a single toolflow.
By parametrising all aspects of the hardware and automating the design space exploration, outstanding performance is achieved across a multitude of networks, targeting more than any existing accelerator.

\begin{figure}
	\centering
	\includegraphics[width=\columnwidth,keepaspectratio]{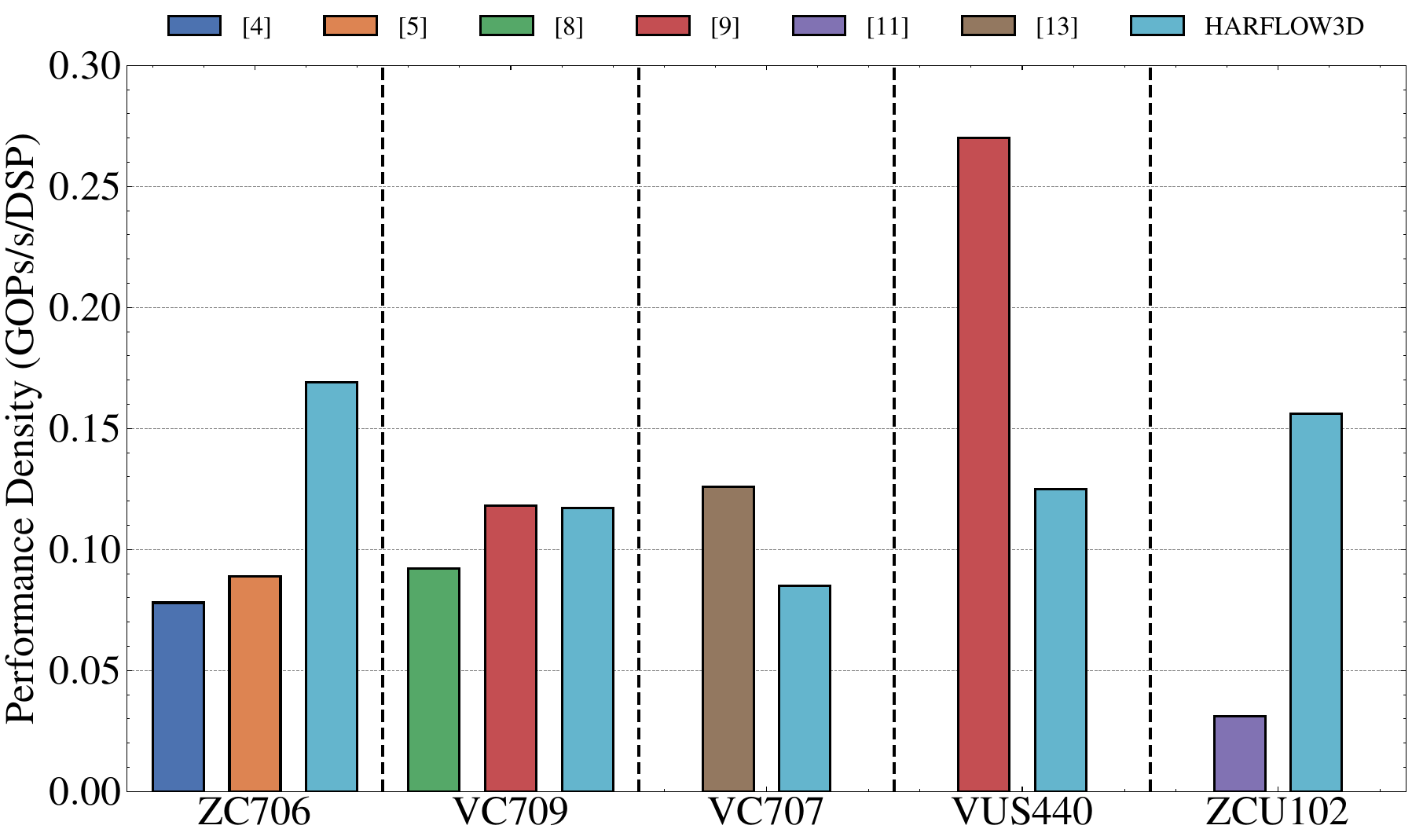}
	\caption{
	   Comparison of the DSP efficiency between HARFLOW3D and previous works. The wide range of the supported FPGA platforms allows for a direct comparison. We demonstrate competitive, and in some cases superior performance.
	}
	\label{fig:c3d_comparison_gops_dsp}
\end{figure}

Figure \ref{fig:c3d_comparison_gops_dsp} provides a more extensive and direct comparison to existing works on C3D model. Since HARFLOW3D can target all of the platforms that previous studies have targeted for C3D, results for each of them have been collected for a direct comparison with all of the existing works. The results are evaluated in DSP efficiency, GOPs normalised over the device's DSPs. As shown in Figure \ref{fig:c3d_comparison_gops_dsp}, HARFLOW3D achieves $1.89\times$ increased DSP efficiency on ZC706 compared to H. Fan\cite{Fan2018}. On ZCU102, $5.03\times$ greater results are achieved than M. Sun\cite{Sun20203DPruning}. On VC709 HARFLOW3D achieves $1.27\times$ better DSP efficiency than Z. Liu\cite{Liu2019AFpgas}, whereas obtains nearly equal performance being only $1.008\times$ off compared to J. Shen\cite{Shen2018}. Compared to T. Teng\cite{Teng2020ExplorationAccelerator} on VC707, the DSP efficiency is $1.48\times$ lesser, however the comparison cannot be considered direct since the specific design uses fixed-point 8 arithmetic precision. In comparison to J. Shen\cite{Shen2018} on VUS440, the DSP efficiency is inferior by $2.16\times$.

\begin{table}[h]
\centering
\caption{Comparison of HARFLOW3D against a GPU for the C3D model. We achieve comparable energy efficiency.}
\label{tab:gpu_comparison}
\begin{tabular}{cc|c}
\hline
			& GPU	& FPGA (HARFLOW3D) \\
\hline
Platform	        & RTX 3090	     & ZCU106 \\
Clock Frequency     & 1.7 GHz		 & 200 MHz \\
Precision	        & 32-bit float	 & 16-bit fixed \\
Latency/clip (ms)	& 6.93		     & 182.81 \\
Power (W)			& 234.1		     & 9.44	\\
Energy/clip (J)		& 1.62		     & 1.72	\\
\hline
\end{tabular}%
\end{table}

Table \ref{tab:gpu_comparison} outlines a comparison between an RTX 3090, a server-grade cutting-edge GPU, and a ZCU106, a mid-range FPGA board.
Despite the difference in scale between the two devices, the results for energy/clip demonstrate the efficiency of the HARFLOW3D toolflow.

%% file: Conclusion/conclusion.tex
\section{Conclusion} \label{conclusion}

This paper presents HARFLOW3D, the first 3D CNN FPGA toolflow for 3D CNNs that supports a wide range of models and devices. A series of transformations during optimization, in conjunction with a novel hardware implementation supporting runtime parametrization of the hardware nodes, enabled comparable and even greater performance in comparison to prior hand-tuned designs. Future directions may be the support of additional 3D CNN models with different backbones, such as Inception-like architectures (I3D). Furthermore, expansion into domains other than HAR, such as 3D semantic segmentation, medical imaging (CT and MRI scans), 3D object detection from point clouds and Transformer-based HAR models will be investigated.